# Complementary structural and functional abnormalities to localise epileptogenic tissue

Jonathan J. Horsley[1], Rhys H. Thomas[2], Fahmida A. Chowdhury[3],
Beate Diehl[3], Andrew W. McEvoy[3], Anna Miserocchi[3], Jane de Tisi[3],
Sjoerd B. Vos[3,4,5], Matthew C. Walker[3], Gavin P. Winston[3,6], John S. Duncan[3],
Yujiang Wang[1,2,3] and Peter N. Taylor[*1,2,3]

**Background**: When investigating suitability for epilepsy surgery, people with drug-refractory focal epilepsy may have intracranial EEG (iEEG) electrodes implanted to localise seizure onset. Diffusion-weighted magnetic resonance imaging (dMRI) may be acquired to identify key white matter tracts for surgical avoidance. Here, we investigate whether structural connectivity abnormalities, inferred from dMRI, may be used in conjunction with functional iEEG abnormalities to aid localisation of the epileptogenic zone (EZ), improving surgical outcomes in epilepsy.

**Methods**: We retrospectively investigated data from 43 patients with epilepsy who had surgery following iEEG. Twenty-five patients (58%) were free from disabling seizures (ILAE 1 or 2) at one year. Interictal iEEG functional, and dMRI structural connectivity abnormalities were quantified by comparison to a normative map and healthy controls. We explored whether the resection of maximal abnormalities related to improved surgical outcomes, in both modalities individually and concurrently. Additionally, we suggest how connectivity abnormalities may inform the placement of iEEG electrodes pre-surgically using a patient case study.

**Findings**: Seizure freedom was 15 times more likely in patients with resection of maximal connectivity and iEEG abnormalities (p=0.008). Both modalities separately distinguished patient surgical outcome groups and when used simultaneously, a decision tree correctly separated 36 of 43 (84%) patients.

**Interpretation**: Our results suggest that both connectivity and iEEG abnormalities may localise epileptogenic tissue, and that these two modalities may provide complementary information in pre-surgical evaluations.

**Funding**: This research was funded by UKRI, CDT in Cloud Computing for Big Data, NIH, MRC, Wellcome Trust and Epilepsy Research UK.

1. CNNP Lab (www.cnnp-lab.com), Interdisciplinary Computing and Complex BioSystems Group, School of Computing, Newcastle University, Newcastle upon Tyne, United Kingdom

2. Translational and Clinical Research Institute, Faculty of Medical Sciences, Newcastle University, Newcastle upon Tyne, United Kingdom

3. Department of Clinical and Experimental Epilepsy, UCL Queen Square Institute of Neurology, University College London, London, United Kingdom

4. Centre for Microscopy, Characterisation, and Analysis, The University of Western Australia, Nedlands, Australia

5. Centre for Medical Image Computing, Computer Science Department, University College London, London, United Kingdom

6. Division of Neurology, Department of Medicine, Queen's University, Kingston, Canada

* Peter.Taylor@newcastle.ac.uk

## Research in context

### Evidence before this study

Intracranial electroencephalography (iEEG) is often used as part of the pre-surgical evaluation for epilepsy surgery. Prior studies have shown inter-ictal data may be able to localise epileptogenic tissue and predict patient surgical outcome. However, some patient outcomes were incorrectly predicted, possibly due to incomplete coverage of the epileptogenic zone with iEEG electrodes. Additionally, diffusion-weighted MRI (dMRI) has frequently highlighted brain abnormalities in people with epilepsy. These abnormalities are not currently used for localisation of epileptogenic tissue in clinical settings.

### Added value of this study

In this work, we showed that dMRI abnormalities may localise epileptogenic regions in patients who subsequently had iEEG implantation and surgery. This is particularly important because existing non-invasive methods were unable to provide a definitive localisation in these individuals. We validated our results using post-surgical outcome scores. Additionally, we showed that within-patients structural dMRI and functional iEEG abnormalities do not necessarily occur in the same regions. This finding suggests that the two modalities may provide complementary information for pre-surgical evaluations. Resection of both the maximal dMRI and iEEG abnormalities was a strong predictor of post-surgical seizure freedom.

### Implications of all the available evidence

Structural dMRI could be used in pre-surgical evaluations, particularly when localisation of the epileptogenic zone is uncertain and iEEG implantation is being considered. Regions with the greatest structural connectivity reductions should be strongly considered for sampling by iEEG electrodes. Our approach allows for the proposal of a personalised iEEG implantation and resection which may lead to improved surgical outcome for an individual patient.

## Introduction

Resective surgery is an effective treatment option for people with drug-resistant focal epilepsy [1]. The target for surgery is the epileptogenic zone (EZ), the part of the brain thought to be responsible for seizure generation [2]. However, localisation of the EZ can be difficult, particularly in patients without visible lesions on MRI.

MRI-negative patients typically have a lower chance of seizure freedom following surgery [3]. To improve localisation of the EZ before surgery, some patients may undergo intracranial EEG (iEEG) implantation [2]. Each additional intracranial electrode accumulates a small but tangible risk to the patient and so there is a finite limit as to the coverage that can be achieved. This means that despite iEEG, there is inherent uncertainty in where to implant and subsequently resect. Improved methods for the localisation, and subsequent resection, of the EZ could improve the rates of seizure freedom following surgery.

The essence of pre-surgical evaluation is the synthesis of a range of data, of varying sources and quality [4]. Newer quantitative methods can assist the traditional qualitative approaches used clinically, and mitigate against unhelpful human biases. Using quantitative techniques, patients with epilepsy may have abnormalities detectable by different modalities, including MRI [5–8], EEG [9–11], MEG [12,13], and diffusion-weighted MRI (dMRI) [14–18]. The quantity, magnitude and location of these abnormalities have been shown to relate to surgical outcome [10,11,19–21]. In addition, different modalities may provide complementary information, such that multimodal analysis can offer an improvement over a single modality [22]. As a result, quantitative methods to incorporate multiple modalities may be able to improve our understanding of seizures, epilepsy and the reasons for surgical failure.

Electrical recordings of brain activity have long been used to identify brain regions implicated in seizure generation. This identification typically involves locating seizure onset regions from ictal data. More recently, normative maps of healthy brain activity have been created using interictal iEEG recordings [10,11,23]. These maps allow for the identification of abnormalities in individual patients by comparing each patient to a normative map. Hypothesising that abnormalities may be epileptogenic, studies have shown that resection of the more abnormal regions related to a better post-surgical outcome [10,11]. These findings suggest that interictal iEEG abnormalities may be able to localise epileptogenic tissue.

Clinically, dMRI is often acquired to surgically avoid key white matter connections [4], rather than to localise abnormal regions of the brain. Diffusion-weighted MRI is used to infer the amount of restriction experienced by water molecules in a given location (connection) of the brain. In healthy white matter, diffusion is anisotropic and is less restricted parallel to the direction of the tract. Diffusion properties (e.g. fractional anisotropy) are used to quantify structural connectivity between regions and are often abnormal in patients with epilepsy [14–18]. Additionally, there is evidence that resection of structural connectivity abnormalities is associated with a better surgical outcome [19,21,24]. Epilepsy is now considered to be a network disorder [25] and connectivity abnormalities may therefore be a biomarker of epileptogenic region(s). As a result, incorporating structural connectivity abnormalities into the pre-surgical evaluation may have the potential to improve surgical outcomes.

In this paper, we investigate how structural connectivity abnormalities and iEEG abnormalities may be used to aid localisation and resection of the EZ in a retrospective cohort of 43 individuals with refractory epilepsy. Specifically, we explore whether:

1. resection of maximal abnormalities in *both* modalities simultaneously is associated with better surgical outcome.

2. both modalities are separately able to distinguish patient outcomes.

3. connectivity abnormalities can be used to guide iEEG implantation.

## Methods

### Patient cohort

We retrospectively studied 43 patients with refractory focal epilepsy from the National Hospital of Neurology and Neurosurgery, London, United Kingdom. Patients were included if they had available preop dMRI, T1w MRI, and iEEG data, additional to postoperative MRI and follow up information. The duration of epilepsy ranged from 5.7 years to 48.3 years (median = 20.2 years, IQR = 10.1 years), and 18 (42%) patients were female. All patients underwent anatomical T1-weighted MRI, diffusion-weighted MRI and iEEG implantation. Of 43 patients, 21 had surgery on the left hemisphere. Of these patients, 61% underwent resection of the temporal lobe, 28% frontal lobe, 7% parietal lobe, 2% combined occipital and parietal lobes and 2% combined temporal and occipital lobes. Post-surgical outcome was assessed using the ILAE classification scale. An ILAE 1 classification indicates complete seizure freedom in a patient, ILAE 2 indicates only auras and ILAE 3+ indicates varying levels of recurring seizures [26]. Good post-surgical outcomes (ILAE 1 or 2) were observed in 58% of patients, with the remainder having poor outcomes (ILAE 3+) at follow-up of 12 months. Outcome did not differ by age, sex, TLE/ETLE diagnosis, side of resection, number of electrodes implanted or presence of a visible lesion on MRI (Table 1). A range of pathologies were present in the patient cohort including focal cortical dysplasia (35%), hippocampal sclerosis (21%), dysembryoplastic neuroepithelial tumors (7%) and cavernoma (5%).

*Table 1: Patient data by post-surgical seizure freedom.*

|  | ILAE 1,2 | ILAE 3+ | Test statistic |
|---|---|---|---|
| n | 25 | 18 |  |
| Age, median (IQR) | 30 (6) | 31.5 (15) | $t = -0.38, p = 0.71$ |
| Sex, male:female | 12:13 | 13:5 | $\chi^2 = 1.63, p = 0.20$ |
| Type, temporal:extratemporal | 16:9 | 10:8 | $\chi^2 = 0.06, p = 0.81$ |
| Side, left:right | 15:10 | 6:12 | $\chi^2 = 2.01, p = 0.16$ |
| Number of electrodes, median (IQR) | 77 (31) | 69.5 (21.25) | $t = 0.57, p = 0.57$ |
| MRI, non-lesional:lesional | 10:15 | 10:8 | $\chi^2 = 0.49, p = 0.48$ |

### dMRI acquisition and processing

Diffusion-weighted MRI acquisition and processing was carried out as described previously [6]. Briefly, the 43 patients and 96 healthy controls were scanned as part of two separate cohorts using different scanning protocols. The first cohort was collected between 2009 and 2013, and had 39 patients and 29 controls. The second cohort was collected between 2014 and 2019, and had 4 patients and 67 controls. Diffusion-weighted MRI data were corrected for signal drift, eddy current and movement artefacts. The b-vectors were then rotated appropriately, before the diffusion data were reconstructed in MNI-152 space using q-space diffeomorphic reconstruction (QSDR). The HCP-1065 tractography atlas was used to determine connections between regions of the Lausanne-60 parcellation scheme, which has 128 cortical and subcortical regions of interest (ROIs). A connection between MNI-152 space regions was defined as present if streamlines connected both regions in the corresponding region pair. A full description of dMRI acquisition and processing is provided in Supplementary Methods 1.

### iEEG acquisition and processing

The 43 patients included here are a subset of a previously studied cohort [10] who also had dMRI. As before, the 21,598 recording contacts from outside the seizure onset and initial propagation zone in the RAM normative cohort of 234 participants with epilepsy were analysed to act as a baseline of presumed non-pathological activity (data collected up to Year 3; http://memory.psych.upenn.edu/RAM). Intracranial EEG acquisition and processing was carried out as described previously [10]. Briefly, 70 seconds of awake interictal recording was extracted for each subject, at least 2 hours away from seizures. After applying a common average reference, the power spectral density in each recording was estimated. The average bandpower was calculated for five frequency bands (delta: 1-4Hz, theta: 4-8Hz, alpha: 8-13Hz, beta: 13-30Hz and gamma: 30-80Hz). Band power estimates were $\log_{10}$ transformed and normalised to sum to 1 for each contact, giving a relative band power. Implanted electrode contacts were assigned to the closest (<5mm) grey matter region of interest according to the Lausanne-60 parcellation scheme. The regional relative band power (for each frequency band) was calculated by taking the mean of electrode contacts assigned to that region. A full description of iEEG acquisition and processing is provided in Supplementary Methods 2.

In this analysis, we do not investigate interictal spikes since this was not predictive of outcome in our previous analysis, and only marginally impacts abnormalities (see figures 5 and S1,S2,S3 in [10]).

### Resection delineation

Since each patient had both pre-operative and post-operative T1-weighted MRI, we were able to quantify which regions had been resected. This was done by linearly registering the post-operative T1w scan to the pre-operative scan and manually delineating the resected tissue as a mask as described previously [27,28]. Using the Lausanne-60 anatomical parcellation, each region within a patient was considered resected if there was a >10% reduction in regional volume post-operatively. The results presented here are robust to differing thresholds (25% and 50%) for considering a region resected (Supplementary Analysis 3).

## Analysis

All data processing was performed using R version 4.12 (https://www.r-project.org), unless otherwise stated.

### Connectivity abnormality calculation

The pipeline for calculating connectivity abnormalities is summarised in Figure 1 panels A-D. For each subject, weighted connectivity matrices were inferred in DSI Studio using fractional anisotropy (FA). ComBat was applied to account for systematic differences in connection weights due to scanner effects [29]. Across subjects, connection weights were corrected for age and sex effects using a robust linear model applied to healthy controls. For each connection i, we calculated the mean $\mu_i$ and standard deviation $\sigma_i$ of connection weights in healthy controls. Connection abnormalities, $A_{ij}$ for each connection i within each patient j, were calculated from the connection strength ($C_{ij}$) using z-scoring:

$$A_{ij} = \frac{C_{ij} - \mu_i}{\sigma_i}$$

To summarise connection abnormalities at a regional level (according to the Lausanne-60 parcellation), we defined the regional connection abnormality, $R_{kj}$ as the mean connection abnormality of all n connections from a given region:

$$R_{kj} = \sum_{i=1}^{n} \frac{A_{ij}}{n}$$

Thus, for each region k in each patient j, we derived a quantitative measure of the abnormality of that regions' white matter connections.

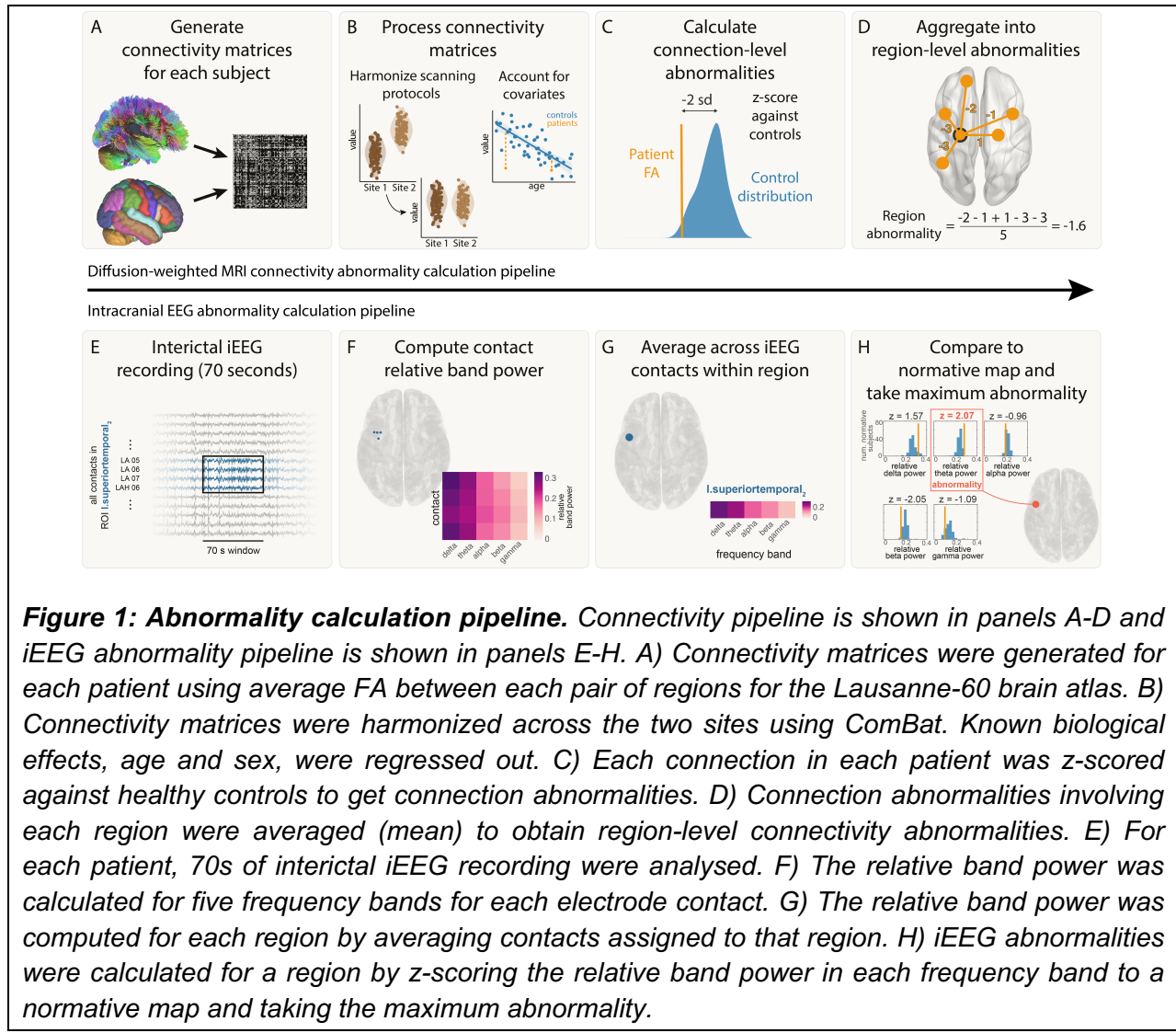

*Figure 1: Abnormality calculation pipeline. Connectivity pipeline is shown in panels A-D and iEEG abnormality pipeline is shown in panels E-H. A) Connectivity matrices were generated for each patient using average FA between each pair of regions for the Lausanne-60 brain atlas. B) Connectivity matrices were harmonized across the two sites using ComBat. Known biological effects, age and sex, were regressed out. C) Each connection in each patient was z-scored against healthy controls to get connection abnormalities. D) Connection abnormalities involving each region were averaged (mean) to obtain region-level connectivity abnormalities. E) For each patient, 70s of interictal iEEG recording were analysed. F) The relative band power was calculated for five frequency bands for each electrode contact. G) The relative band power was computed for each region by averaging contacts assigned to that region. H) iEEG abnormalities were calculated for a region by z-scoring the relative band power in each frequency band to a normative map and taking the maximum abnormality.*

### Intracranial EEG abnormality calculation

The pipeline for calculating connectivity abnormalities is summarised in Figure 1 panels E-H. For each frequency band, f, and region, k, within each patient j, iEEG abnormalities were calculated using z-scoring:

$$z_{fkj} = \frac{b_{fkj} - \mu_{fk}}{\sigma_{fk}},$$

where $b_{fkj}$ was the band power for a given frequency band (f), region (k) and patient (j). Further, $\mu_{fk}$ and $\sigma_{fk}$ were the mean and standard deviation across all patients for a given frequency band and region. We then defined the patient's band power abnormality for each ROI and time window as the maximum absolute z-score across the five frequency bands:

$$B_{kj} = max_f(|z_{fkj}|)$$

This maximum absolute z-score was previously shown to not be systematically derived from a particular frequency band (see figures S10,S11 in [10]).

### Relating connectivity and iEEG abnormalities

For each patient, we calculated connectivity and iEEG abnormalities in each region. In this analysis, we took the subset of regions with iEEG implantation, i.e. regions with both connectivity and iEEG abnormalities. Two example patients are shown in Figure 2. Next, we combined the abnormalities and used a support vector machine (SVM) to separate the abnormalities into spared and resected zones. If the SVM successfully split a patient's 2D abnormality space into two different (spared and resected) zones, then we defined this as successful separation. This separability was not possible in some patients whose resected and spared abnormalities were overlapping. This SVM approach was used to determine whether a region with the greatest abnormality in both modalities (i.e. maximal) would likely be resected or spared within that patient. We tested whether this tendency to resect maximal abnormalities related to surgical outcome using a chi-squared test and odds ratios.

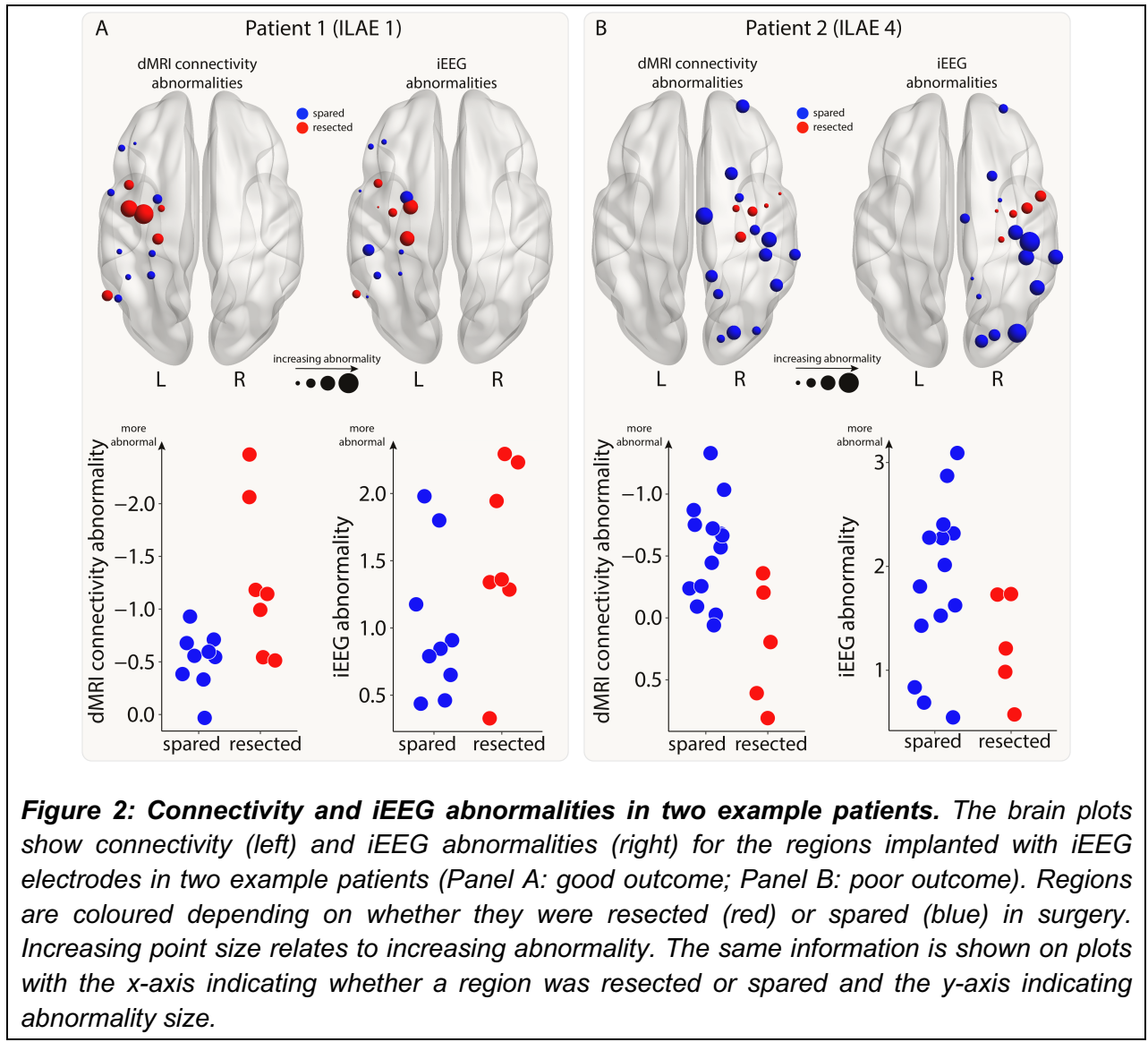

*Figure 2: Connectivity and iEEG abnormalities in two example patients. The brain plots show connectivity (left) and iEEG abnormalities (right) for the regions implanted with iEEG electrodes in two example patients (Panel A: good outcome; Panel B: poor outcome). Regions are coloured depending on whether they were resected (red) or spared (blue) in surgery. Increasing point size relates to increasing abnormality. The same information is shown on plots with the x-axis indicating whether a region was resected or spared and the y-axis indicating abnormality size.*

### Computing $D_{RS}$

We defined a statistic, $D_{RS}$, as the distinguishability between resected and spared tissue [10,12], equivalent to the normalized Mann-Whitney U-statistic. This measures the extent to which abnormalities occur in the spared regions compared to the resected regions. In the same way as a (ROC) AUC, $D_{RS}$ is a value between 0 and 1, where 0 indicates the largest abnormalities are all in the resected regions, and vice versa. As a result, we might expect a seizure-freedom in patients with a $D_{RS}$ value of close to 0, since the most abnormal tissue is removed. For each patient, $D_{RS}$ values were computed separately for connectivity and iEEG abnormalities. For connectivity $D_{RS}$, all brain regions were used in the calculation. This calculation differed from iEEG $D_{RS}$, which only used the regions covered with iEEG implantation. We tested whether there was a statistical difference between surgical outcome groups using Mann-Whitney U tests and quantified this difference using the area under the curve (AUC) of receiver operating characteristic curves.

### Relating connectivity and iEEG $D_{RS}$ values

To classify patients based on their surgical outcome, we fitted a decision tree. This decision tree took connectivity and iEEG $D_{RS}$ values as an input and divided the space (2-dimensional scatter plot) into seizure-free and non-seizure-free zones. We ensured that the decision tree performed one cut based on iEEG abnormalities and another cut based on connectivity abnormalities, but did not specify the order or placement of these cuts. We then counted the number of patients correctly classified to give a classification accuracy. To assess how our approach might perform on new, unseen patients, we performed an alternative approach using leave-one-out cross validation (Supplementary Analysis 2).

### Ethics

The study was approved by the National Hospital for Neurology and Neurosurgery and the Institute of Neurology Joint Research Ethics Committee. Pseudonymised data were analysed under the approval of the Newcastle University Ethics Committee (2225/2017).

### Role of funders

None of the funding sources played a role in the study design, data collection, data analyses, interpretation, or writing the manuscript.

# Results

## Resection of maximal abnormalities was associated with better surgical outcomes

First, we investigated whether resection of maximal abnormalities was associated with better surgical outcomes. We illustrate our findings with two example patients (Figure 3A). In our first patient, the SVM separated the resected and spared regions, in terms of abnormality (left panel of Figure 3A). Abnormalities were not necessarily concordant. For example one resected region is highly abnormal in dMRI, but not in iEEG data (indicated by red single arrow). Furthermore, two other regions were abnormal for iEEG, but not dMRI data (red double arrow). Although abnormalities were not concordant, the SVM did separate the resected and spared regions. This patient had abnormalities resected (i.e. red shading in upper right in contrast to lower left) and was seizure-free. We term this successful separation *and* resection as a 'tendency to resect maximal abnormality'. Furthermore, the lack of full concordance between modalities in patient 1 suggests complementary information across modalities.

In our second patient (right panel of Figure 3A), regions which were spared, shown in blue, were abnormal in both modalities (iEEG on the x-axis, dMRI on the y-axis). In contrast, the resected regions were much more normal (red data points closer to the axes origin). Furthermore, the support vector machine separated the two (resected and spared) groups well (shaded areas of graph). Taken together, the abnormalities across the two modalities could be clearly separated by their resection, but those regions which were resected were not abnormal in either modality. In contrast to patient 1, this patient was subsequently not seizure-free (ILAE 4).

We next applied a SVM to all patients to separate regions into spared and resected zones, based on abnormalities in the two modalities. Of the 43 patients in our cohort, 28 had a clear separation of resected and spared regions using the SVM. Patients with a tendency to have maximal abnormalities resected (i.e. those similar to patient 2) were 15 times more likely to be seizure-free than those that had maximal abnormalities spared (odds ratio 95% confidence interval = [2.26, 99.64], Chi-squared p-value = 0.008, Figure 3B). For those patients with no clear separation of resected and spared regions, neither seizure-free nor not-seizure-free outcomes were more likely (seizure-free: n=11, not-seizure-free: n=4; p=0.12).

In a supplementary analysis, we found that there was no cohort-wide correlation between iEEG and connectivity abnormalities (Supplementary Analysis 1). This finding, along with those in figure 3A, suggests complementary information from the two modalities. Results were consistent across subgroups of both MRI-positive (n=23) and MRI-negative (n=20) patients (Supplementary Analysis 6).

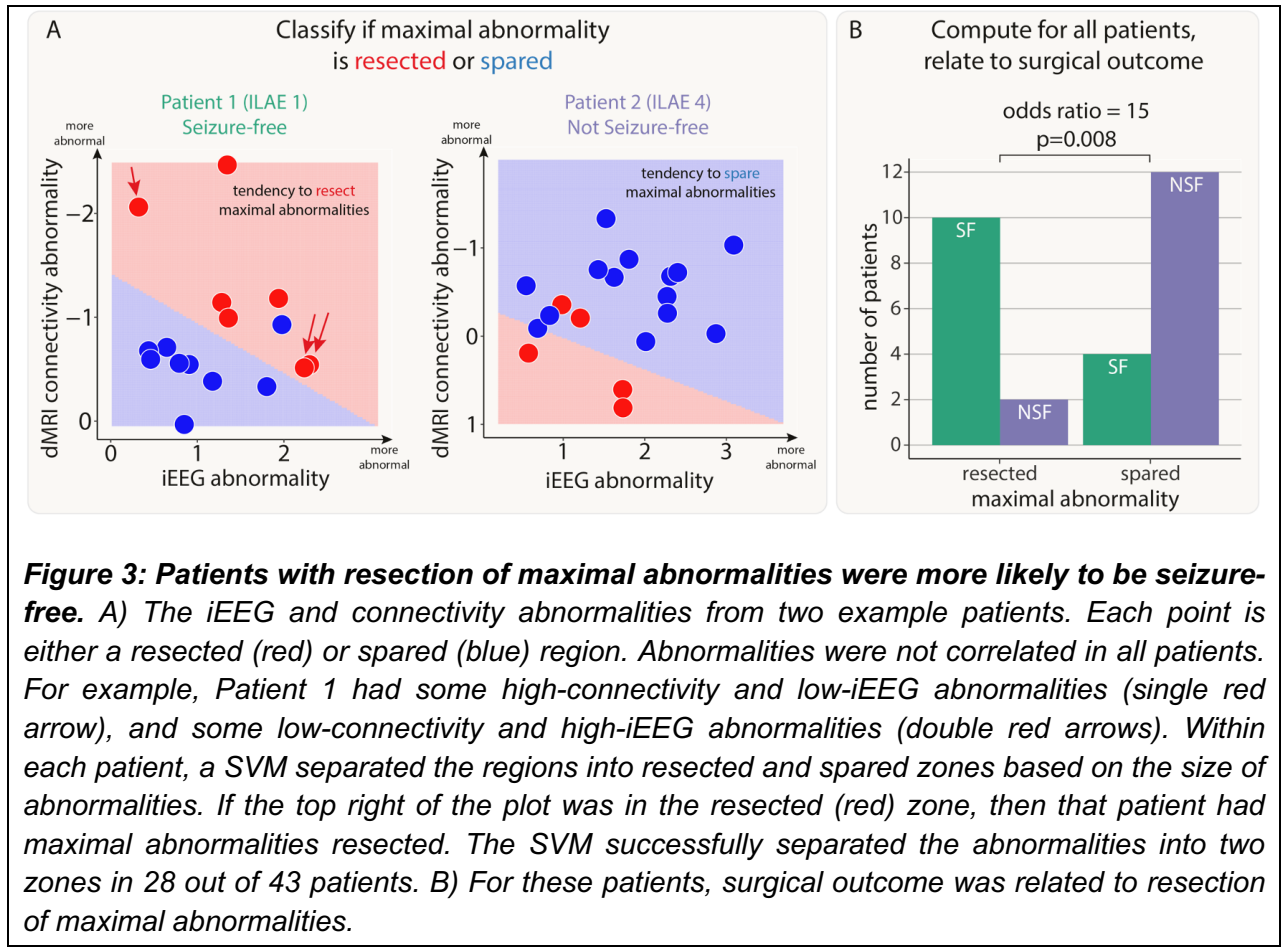

*Figure 3: Patients with resection of maximal abnormalities were more likely to be seizure-free.* A) The iEEG and connectivity abnormalities from two example patients. Each point is either a resected (red) or spared (blue) region. Abnormalities were not correlated in all patients. For example, Patient 1 had some high-connectivity and low-iEEG abnormalities (single red arrow), and some low-connectivity and high-iEEG abnormalities (double red arrows). Within each patient, a SVM separated the regions into resected and spared zones based on the size of abnormalities. If the top right of the plot was in the resected (red) zone, then that patient had maximal abnormalities resected. The SVM successfully separated the abnormalities into two zones in 28 out of 43 patients. B) For these patients, surgical outcome was related to resection of maximal abnormalities.

## Both modalities distinguish patient outcomes and provide complementary information

Next, we investigated whether resection of the largest abnormalities in both modalities could separately distinguish patient surgical outcome in the full cohort of patients. We analysed abnormalities in the resected and spared regions using the $D_{RS}$ measure. Applying $D_{RS}$ individually to connectivity abnormalities (AUC = 0.75, p = 0.003; Figure 4A) and iEEG abnormalities (AUC = 0.67, p = 0.03; Figure 4B) separated outcome groups well. Results were consistent across subgroups of both MRI-positive (n=23) and MRI-negative (n=20) patients (Supplementary Analysis 6).

Perhaps unsurprisingly given the lack of correlation in the underlying abnormalities (Figure 3A, Figure S1) and the fact that the connectivity measure included all regions, $D_{RS}$ values in both modalities were uncorrelated across patients (r = 0.03, p = 0.84). Since both iEEG and connectivity abnormalities were separately predictive of patient outcome, but the underlying abnormalities were typically uncorrelated, useful complementary information may exist when combining the two modalities.

We applied a decision tree to dMRI and iEEG $D_{RS}$ values simultaneously to classify patients as seizure-free or non-seizure-free. Using the full cohort, 36 out of 43 patients (84%) were correctly classified (Figure 4C). We compared to an alternative approach to predict surgical outcome of unseen patients using leave-one-out cross validation (Supplementary Analysis 2). This alternative approach predicted patient outcomes with an accuracy of 72% (sensitivity = 0.76, specificity = 0.67).

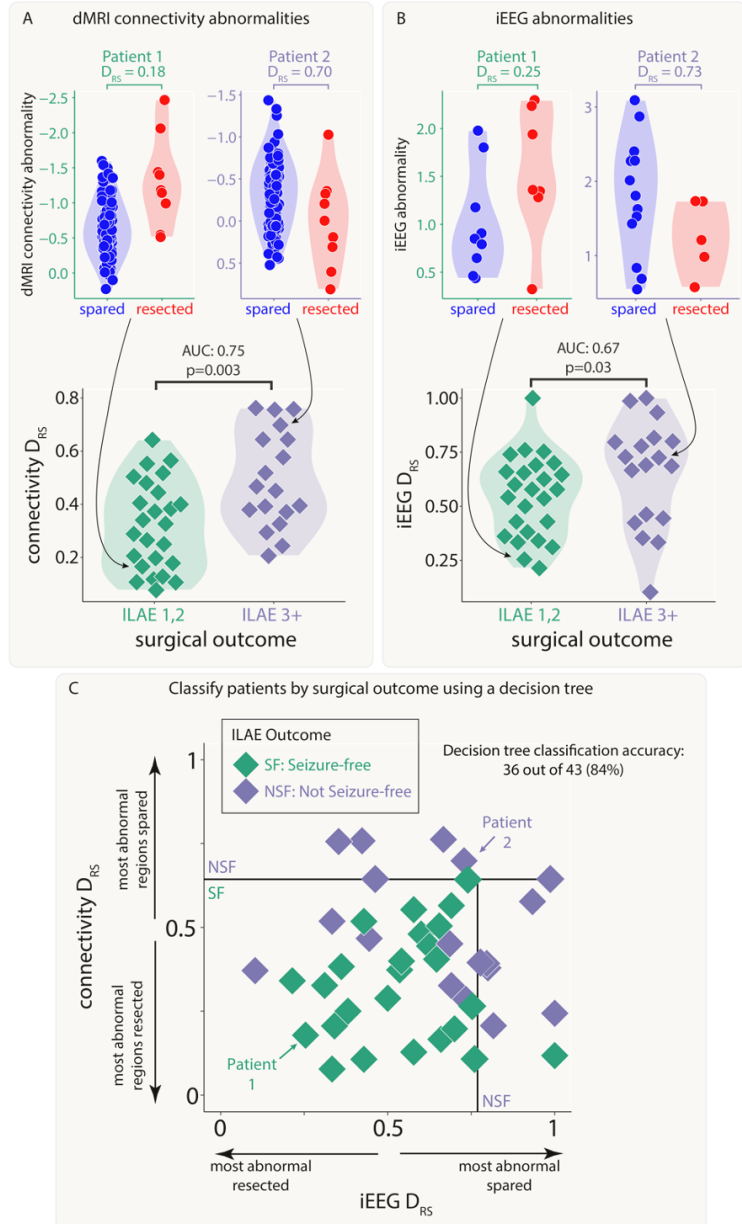

*Figure 4: Connectivity and iEEG abnormality distribution in resected versus spared tissue explains post-surgical seizure freedom.* Both A) connectivity $D_{RS}$ and B) iEEG $D_{RS}$ were used to separate patients based on surgical outcome. The top plots in each panel show regional abnormalities, indicated with circular points, in example patients. The bottom plots in each panel show patient $D_{RS}$ values, indicated with diamond points. C) A decision tree was fit to both $D_{RS}$ values simultaneously to classify patient outcome, achieving an accuracy of 84%.

### dMRI abnormalities may inform iEEG placement and surgical resection

We next retrospectively investigated the feasibility of using dMRI to inform iEEG placement and subsequent surgical resection. We present a case study using a single patient (Figure 5).

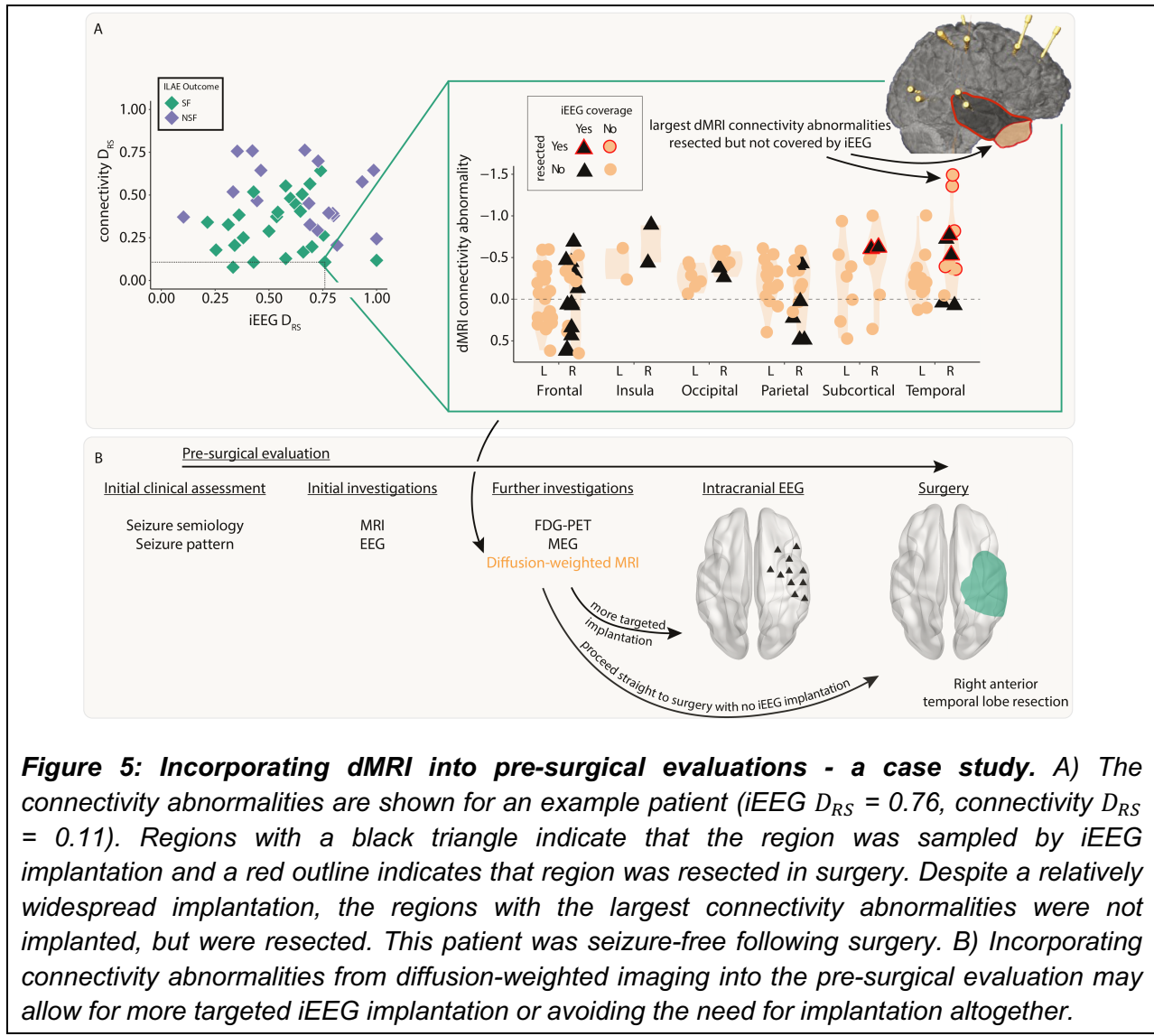

*Figure 5: Incorporating dMRI into pre-surgical evaluations - a case study. A) The connectivity abnormalities are shown for an example patient (iEEG $D_{RS}$ = 0.76, connectivity $D_{RS}$ = 0.11). Regions with a black triangle indicate that the region was sampled by iEEG implantation and a red outline indicates that region was resected in surgery. Despite a relatively widespread implantation, the regions with the largest connectivity abnormalities were not implanted, but were resected. This patient was seizure-free following surgery. B) Incorporating connectivity abnormalities from diffusion-weighted imaging into the pre-surgical evaluation may allow for more targeted iEEG implantation or avoiding the need for implantation altogether.*

This patient underwent a non-invasive pre-surgical evaluation that was inconclusive (Figure 5). This evaluation included semiology, scalp video-EEG, MRI, FDG-PET and MEG. Sufficient uncertainty surrounding the location of the EZ remained, so the patient underwent a large intracranial EEG implantation in the right hemisphere. This implantation included frontal, insula, parietal, temporal, subcortical and occipital regions. The patient proceeded to a right anterior temporal lobe resection and was seizure free post-operatively (ILAE 1).

Pre-operative dMRI was also acquired for this patient. Our retrospective analysis presented here suggests the right temporal pole and right inferior temporal gyrus had the greatest abnormalities compared to what would be expected in health (Figure 5A, top right inset orange circled red anterior area). These regions were not implanted with iEEG electrodes and therefore the iEEG $D_{RS}$ analysis unsurprisingly performed poorly in this patient (iEEG $D_{RS}$ = 0.76). The connectivity analysis correctly predicted that the patient would be seizure-free following surgery (dMRI $D_{RS}$ = 0.11). Four additional case study patients are presented in Supplementary Analysis 4. Incorporating the analysis of dMRI abnormalities into the pre-surgical evaluation process (Figure 5B) may have suggested a modified implantation or to more anterior temporal regions, or even to proceed straight to surgery.

## Discussion

Structural connectivity abnormalities and iEEG abnormalities may be used to predict post-surgical outcome following epilepsy surgery. Specifically, we found that both modalities were separately able to distinguish outcome groups. When used together, iEEG and connectivity abnormalities provided good predictive ability and complementary information. Additionally, incorporating dMRI abnormalities into pre-surgical evaluations could aid placement of iEEG electrodes and subsequent resection.

Whilst traditionally used to avoid resection of key white matter tracts, dMRI may also be beneficial for the localisation of epileptogenic tissue. Few studies have evaluated structural connectivity abnormalities in this context. Fewer abnormalities of individual ipsilateral connections [24] and a smaller number of regional connectivity abnormalities remaining after surgery [19] have both been shown to be associated with better surgical outcomes in TLE [21]. Our approach differed from these studies. We analysed a more heterogenous cohort, including extra-temporal cases, and considered abnormalities specifically as FA reductions, since they are more often observed than FA increases in white matter connections in patients with epilepsy [14]. Nevertheless, we also found that the surgical removal of larger connectivity abnormalities in individual patients was indicative of seizure freedom. These results suggest that FA reductions may have potential as a localising biomarker of epileptogenic tissue in extratemporal epilepsy as well as TLE.

Intracranial EEG recordings are primarily acquired to identify seizure onset regions, rather than for interpretation of interictal activity, although the latter may be informative [30–34]. Given the invasive nature of iEEG recordings, it is imperative to extract maximum value from the data, especially since many patients continue to have seizures post-operatively. Our approach uses two data types (dMRI, interictal iEEG) which are commonly acquired, but traditionally assigned lower clinical importance for localisation. Our quantitative demonstration that both modalities have complementary localisation information suggests these may be clinically useful. Compared to dMRI, the sampling from iEEG is more sparse and non-uniform across patients. This non-uniformity is exaggerated by the heterogeneous cohort of temporal and extratemporal patients. The ability of our SVM approach and iEEG $D_{RS}$ measure to detect epileptogenic regions is clearly dependent on those regions being covered by electrode implantation (in addition to some non-pathological tissue). Given that regions targeted with iEEG implantation are inherently suspected as epileptogenic from other biomarkers of epileptogenicity (semiology, ictal onset, MRI abnormalities etc.), it is likely but not guaranteed that the implantation covers the epileptogenic zone. Diffusion-weighted MRI has full coverage of the brain so does not suffer from this limitation. As a result, dMRI may prove a useful additional modality to consider before iEEG implantation, particularly when localisation from other modalities is inconclusive. Perhaps surprisingly, we found little evidence of abnormalities in both modalities occurring in the same regions within patients. Whilst there was a correlation in some patients (e.g. Patient 1 in Figures 2 and 3), this was not consistently observed across all patients (Supplementary Analysis 1). Connectivity abnormalities and iEEG abnormalities may therefore be driven by different underlying mechanisms in a distributed epileptogenic network, possibly involving excitotoxicity [35], ischaemia [36] or protein aggregation [37], which may or may not manifest in abnormal neural dynamics. Further, seizures could be associated with different mechanisms both within and across patients, with localised abnormal dynamics at seizure onset regions affecting more widespread connectivity as seizures spread. Hence, both modalities can provide complementary information in identifying epileptogenic regions to be targeted by resective surgery.

The importance of quantitatively analysing multimodal data is increasingly being recognised in epilepsy research [38–41]. Our approach of quantifying abnormalities relative to a normative dataset could be easily extended to more than two modalities such as MEG [12], T1-derived structural abnormalities [42,43], or fMRI-derived functional abnormalities [44,45]. Epileptogenic regions may or may not be measurably abnormal across several different modalities. Further research is needed to learn when to expect concordance or discordance, as this may be specific to the type of pathology. Machine learning approaches such as SVM and decision trees, as applied here, could similarly be used to identify which patients had abnormalities resected across three or more modalities. In our work, we wanted to investigate the predictive ability of modalities not usually considered in pre-surgical evaluations. In reality, this would be used in conjunction with traditional clinical parameters. This incorporation of additional data may further improve the retrospective (and eventually prospective) prediction of surgical outcome. Our approach retrospectively analysed patients to predict seizure freedom. Whilst this is important, it is clearly more clinically useful to prospectively predict seizure freedom following surgery. The methods that we use in this study could be modified for prospective use. Both dMRI and iEEG abnormalities could be calculated pre-surgery. Once a proposed resection is formulated, both the SVM and $D_{RS}$ analyses could be applied using these hypothetical/proposed regions to generate a prediction of whether the patient would become seizure free or not. This may inform resection pre-operatively, obviously notwithstanding for what is actually feasible for resection.

Patients who require iEEG are less likely to be seizure free than are those who do not need this, because of the inherent selection bias towards those in whom the location of the epileptogenic zone and network are uncertain [46]. The development of methods that can accurately determine the epileptogenic zone is therefore particularly important for these patients. The results presented here are promising given the heterogeneous cohort of patients with iEEG implantation, achieving comparable accuracies with other studies [47–49]. Our approach for detecting abnormalities in interictal iEEG activity involved taking the maximum relative bandpower abnormality across five frequency bands [10]. This was done to account for various types of interictal abnormalities that could occur. Employing other approaches, including analysing specific frequency bands, may also be beneficial. The fact that our approach is entirely data driven is a strength of our study without the need to pre-define what is or is not epileptiform. However, there are limitations to this study. Firstly, our sample size is relatively small and from a single site. Replication using larger, multi-site, cohorts will be important for future translation [50]. This replication should be achievable since normative mapping approaches have been applied by other sites using both iEEG [11] and dMRI [24]. Secondly, our method to calculate a region's connectivity abnormality averages the abnormality in all white matter connections to/from that region. This implicitly assumes that connection abnormalities behave similarly, which may not necessarily be the case.

Future work in this area could aim to streamline and quantify the pre-surgical evaluation process. In particular, it could investigate whether an individual patient a) could proceed straight to surgery without the need for iEEG implantation if dMRI abnormalities concur with other modalities, b) requires a targeted iEEG implantation, informed by dMRI and other modalities or c) is unlikely to be a good surgical candidate if abnormalities are too widespread. Taken together, our results suggest that it is possible to determine the surgical outcome of patients with a good degree of accuracy using both dMRI and iEEG abnormalities. Incorporating this information into pre-surgical evaluations may increase the likelihood of seizure freedom for those patients whose epilepsy was previously difficult to localise.

## Contributors

Conception and design: JJH, PNT; Acquisition and analysis of data: FAC, BD, AWM, AM, JdT, SBV, MCW, GPW, JSD, YW, PNT; Formal analysis: JJH, PNT; Supervision: RHT, PNT; Writing - original draft: JJH, PNT; Writing - review and editing: JJH, RHT, SBV, MCW, GPW, JSD, YW, PNT.

## Declaration of interests

Nothing to report.

## Acknowledgements

The authors acknowledge the facilities and scientific and technical assistance of the National Imaging Facility, a National Collaborative Research Infrastructure Strategy (NCRIS) capability, at the Centre for Microscopy, Characterisation, and Analysis, the University of Western Australia. B.D. receives support from the NIH National Institute of Neurological Disorders and Stroke U01-NS090407 (Center for SUDEP Research) and Epilepsy Research UK. Y.W. gratefully acknowledges funding from Wellcome Trust (208940/Z/17/Z) and is supported by a UKRI Future Leaders Fellowship (MR/V026569/1). G.P.W. was supported by the MRC (G0802012, MR/M00841X/1). P.N.T. is supported by a UKRI Future Leaders Fellowship (MR/T04294X/1). J.J.H. and T.W.O are supported by the Centre for Doctoral Training in Cloud Computing for Big Data (EP/L015358/1). JD is grateful to Wellcome Trust (WT106882) and Epilepsy Research UK. We are grateful to the Epilepsy Society for supporting the Epilepsy Society MRI scanner. This work was supported by the National Institute for Health Research University College London Hospitals Biomedical Research Centre.

## Data sharing statement

Data to reproduce the main findings will be made available upon request.

# References

[1] Wiebe S. Brain surgery for epilepsy. The Lancet 2003;362:s48–9. https://doi.org/10.1016/S0140-6736(03)15075-1.

[2] Rosenow F, Luders H. Presurgical evaluation of epilepsy. Brain 2001;124:1683–700. https://doi.org/10.1093/brain/124.9.1683.

[3] Bell ML, Rao S, So EL, Trenerry M, Kazemi N, Matt Stead S, et al. Epilepsy surgery outcomes in temporal lobe epilepsy with a normal MRI. Epilepsia 2009;50:2053–60. https://doi.org/10.1111/j.1528-1167.2009.02079.x.

[4] Duncan JS, Winston GP, Koepp MJ, Ourselin S. Brain imaging in the assessment for epilepsy surgery. The Lancet Neurology 2016;15:420–33. https://doi.org/10.1016/S1474-4422(15)00383-X.

[5] Whelan CD, Altmann A, Botía JA, Jahanshad N, Hibar DP, Absil J, et al. Structural brain abnormalities in the common epilepsies assessed in a worldwide ENIGMA study. Brain 2018;141:391–408. https://doi.org/10.1093/brain/awx341.

[6] Horsley JJ, Schroeder GM, Thomas RH, Tisi J de, Vos SB, Winston GP, et al. Volumetric and structural connectivity abnormalities co-localise in TLE. NeuroImage: Clinical 2022:103105. https://doi.org/10.1016/j.nicl.2022.103105.

[7] Keller SS, Roberts N. Voxel-based morphometry of temporal lobe epilepsy: An introduction and review of the literature. Epilepsia 2008;49:741–57. https://doi.org/10.1111/j.1528-1167.2007.01485.x.

[8] Morgan VL, Conrad BN, Abou-Khalil B, Rogers BP, Kang H. Increasing structural atrophy and functional isolation of the temporal lobe with duration of disease in temporal lobe epilepsy. Epilepsy Research 2015;110:171–8. https://doi.org/10.1016/j.eplepsyres.2014.12.006.

[9] Pillai J, Sperling MR. Interictal EEG and the Diagnosis of Epilepsy. Epilepsia 2006;47:14–22. https://doi.org/10.1111/j.1528-1167.2006.00654.x.

[10] Taylor PN, Papasavvas CA, Owen TW, Schroeder GM, Hutchings FE, Chowdhury FA, et al. Normative brain mapping of interictal intracranial EEG to localize epileptogenic tissue. Brain 2022;145:939–49. https://doi.org/10.1093/brain/awab380.

[11] Bernabei JM, Sinha N, Arnold TC, Conrad E, Ong I, Pattnaik AR, et al. Normative intracranial EEG maps epileptogenic tissues in focal epilepsy. Brain 2022;145:1949–61. https://doi.org/10.1093/brain/awab480.

[12] Owen TW, Schroeder GM, Janiukstyte V, Hall GR, McEvoy A, Miserocchi A, et al. <Span style="font-variant:small-caps;">MEG</span> abnormalities and mechanisms of surgical failure in neocortical epilepsy. Epilepsia 2023:epi.17503. https://doi.org/10.1111/epi.17503.

[13]     Kudo K, Morise H, Ranasinghe KG, Mizuiri D, Bhutada AS, Chen J, et al. Magnetoencephalography Imaging Reveals Abnormal Information Flow in Temporal Lobe Epilepsy. Brain Connectivity 2022;12:362–73. https://doi.org/10.1089/brain.2020.0989.

[14]     Hatton SN, Huynh KH, Bonilha L, Abela E, Alhusaini S, Altmann A, et al. White matter abnormalities across different epilepsy syndromes in adults: An ENIGMA-Epilepsy study. Brain 2020;143:2454–73. https://doi.org/10.1093/brain/awaa200.

[15]     Besson P, Dinkelacker V, Valabregue R, Thivard L, Leclerc X, Baulac M, et al. Structural connectivity differences in left and right temporal lobe epilepsy. NeuroImage 2014;100:135–44. https://doi.org/10.1016/j.neuroimage.2014.04.071.

[16]     Bonilha L, Helpern JA, Sainju R, Nesland T, Edwards JC, Glazier SS, et al. Presurgical connectome and postsurgical seizure control in temporal lobe epilepsy. Neurology 2013;81:1704–10. https://doi.org/10.1212/01.wnl.0000435306.95271.5f.

[17]     Owen TW, Tisi J, Vos SB, Winston GP, Duncan JS, Wang Y, et al. Multivariate white matter alterations are associated with epilepsy duration. European Journal of Neuroscience 2021;53:2788–803. https://doi.org/10.1111/ejn.15055.

[18]     Concha L, Kim H, Bernasconi A, Bernhardt BC, Bernasconi N. Spatial patterns of water diffusion along white matter tracts in temporal lobe epilepsy. Neurology 2012;79:455–62. https://doi.org/10.1212/WNL.0b013e31826170b6.

[19]     Sinha N, Wang Y, Moreira da Silva N, Miserocchi A, McEvoy AW, Tisi J de, et al. Structural brain network abnormalities and the probability of seizure recurrence after epilepsy surgery. Neurology 2020:10.1212/WNL.0000000000011315. https://doi.org/10.1212/WNL.0000000000011315.

[20]     Bonilha L, Keller SS. Quantitative MRI in refractory temporal lobe epilepsy: Relationship with surgical outcomes. Quantitative Imaging in Medicine and Surgery 2015;5.

[21]     Keller SS, Glenn GR, Weber B, Kreilkamp BAK, Jensen JH, Helpern JA, et al. Preoperative automated fibre quantification predicts postoperative seizure outcome in temporal lobe epilepsy. Brain 2017;140:68–82. https://doi.org/10.1093/brain/aww280.

[22]     Duez L, Beniczky S, Tankisi H, Hansen PO, Sidenius P, Sabers A, et al. Added diagnostic value of magnetoencephalography (MEG) in patients suspected for epilepsy, where previous, extensive EEG workup was unrevealing. Clinical Neurophysiology 2016;127:3301–5. https://doi.org/10.1016/j.clinph.2016.08.006.

[23]     Frauscher B, Ellenrieder N von, Zelmann R, Doležalová I, Minotti L, Olivier A, et al. Atlas of the normal intracranial electroencephalogram: Neurophysiological awake activity in different cortical areas. Brain 2018;141:1130–44. https://doi.org/10.1093/brain/awy035.

[24]     Bonilha L, Jensen JH, Baker N, Breedlove J, Nesland T, Lin JJ, et al. The brain connectome as a personalized biomarker of seizure outcomes after temporal lobectomy. Neurology 2015;84:1846–53. https://doi.org/10.1212/WNL.0000000000001548.

[25]    Bernhardt BC, Bonilha L, Gross DW. Network analysis for a network disorder: The emerging role of graph theory in the study of epilepsy 2015:9.

[26]    Durnford AJ, Rodgers W, Kirkham FJ, Mullee MA, Whitney A, Prevett M, et al. Very good inter-rater reliability of Engel and ILAE epilepsy surgery outcome classifications in a series of 76 patients. Seizure 2011;20:809–12. https://doi.org/10.1016/j.seizure.2011.08.004.

[27]    Wang Y, Sinha N, Schroeder GM, Ramaraju S, McEvoy AW, Miserocchi A, et al. Interictal intracranial electroencephalography for predicting surgical success: The importance of space and time. Epilepsia 2020;61:1417–26. https://doi.org/10.1111/epi.16580.

[28]    Taylor PN, Sinha N, Wang Y, Vos SB, Tisi J de, Miserocchi A, et al. The impact of epilepsy surgery on the structural connectome and its relation to outcome. NeuroImage: Clinical 2018;18:202–14. https://doi.org/10.1016/j.nicl.2018.01.028.

[29]    Fortin J-P, Parker D, Tunç B, Watanabe T, Elliott MA, Ruparel K, et al. Harmonization of multi-site diffusion tensor imaging data. NeuroImage 2017;161:149–70. https://doi.org/10.1016/j.neuroimage.2017.08.047.

[30]    Gunnarsdottir KM, Li A, Smith RJ, Kang J-Y, Korzeniewska A, Crone NE, et al. Source-sink connectivity: A novel interictal EEG marker for seizure localization. Brain 2022;145:3901–15. https://doi.org/10.1093/brain/awac300.

[31]    Paulo DL, Wills KE, Johnson GW, Gonzalez HFJ, Rolston JD, Naftel RP, et al. SEEG Functional Connectivity Measures to Identify Epileptogenic Zones: Stability, Medication Influence, and Recording Condition. Neurology 2022:10.1212/WNL.0000000000200386. https://doi.org/10.1212/WNL.0000000000200386.

[32]    Goodale SE, González HFJ, Johnson GW, Gupta K, Rodriguez WJ, Shults R, et al. Resting-State SEEG May Help Localize Epileptogenic Brain Regions. Neurosurgery 2020;86:792–801. https://doi.org/10.1093/neuros/nyz351.

[33]    Zweiphenning W, Klooster MA van 't, Klink NEC van, Leijten FSS, Ferrier CH, Gebbink T, et al. Intraoperative electrocorticography using high-frequency oscillations or spikes to tailor epilepsy surgery in the Netherlands (the HFO trial): A randomised, single-blind, adaptive non-inferiority trial. The Lancet Neurology 2022;21:982–93. https://doi.org/10.1016/S1474-4422(22)00311-8.

[34]    Li A, Huynh C, Fitzgerald Z, Cajigas I, Brusko D, Jagid J, et al. Neural fragility as an EEG marker of the seizure onset zone. Nature Neuroscience 2021;24:1465–74. https://doi.org/10.1038/s41593-021-00901-w.

[35]    Henshall DC. Apoptosis signalling pathways in seizure-induced neuronal death and epilepsy. Biochemical Society Transactions 2007;35:421–3. https://doi.org/10.1042/BST0350421.

[36]	Farrell JS, Wolff MD, Teskey GC. Neurodegeneration and Pathology in Epilepsy: Clinical and Basic Perspectives. In: Beart P, Robinson M, Rattray M, Maragakis NJ, editors. Neurodegenerative Diseases: Pathology, Mechanisms, and Potential Therapeutic Targets, Cham: Springer International Publishing; 2017, p. 317–34. https://doi.org/10.1007/978-3-319-57193-5_12.

[37]	Fricker M, Tolkovsky AM, Borutaite V, Coleman M, Brown GC. Neuronal Cell Death. Physiol Rev 2018;98:68.

[38]	Duncan JS. Epilepsy in the 21st century. The Lancet Neurology 2022;21:501–3. https://doi.org/10.1016/S1474-4422(22)00175-2.

[39]	Proix T, Bartolomei F, Guye M, Jirsa VK. Individual brain structure and modelling predict seizure propagation. Brain 2017:14.

[40]	Makhalova J, Medina Villalon S, Wang H, Giusiano B, Woodman M, Bénar C, et al. Virtual epileptic patient brain modeling: Relationships with seizure onset and surgical outcome. Epilepsia 2022;63:1942–55. https://doi.org/10.1111/epi.17310.

[41]	Shah P, Ashourvan A, Mikhail F, Pines A, Kini L, Oechsel K, et al. Characterizing the role of the structural connectome in seizure dynamics. Brain 2019;142:1955–72. https://doi.org/10.1093/brain/awz125.

[42]	Spitzer H, Ripart M, Whitaker K, D'Arco F, Mankad K, Chen AA, et al. Interpretable surface-based detection of focal cortical dysplasias: A Multi-centre Epilepsy Lesion Detection study. Brain 2022;145:3859–71. https://doi.org/10.1093/brain/awac224.

[43]	Bernhardt BC, Hong S-J, Bernasconi A, Bernasconi N. Magnetic resonance imaging pattern learning in temporal lobe epilepsy: Classification and prognostics: MRI Profiling in TLE. Annals of Neurology 2015;77:436–46. https://doi.org/10.1002/ana.24341.

[44]	Morgan VL, Johnson GW, Cai LY, Landman BA, Schilling KG, Englot DJ, et al. MRI network progression in mesial temporal lobe epilepsy related to healthy brain architecture. Network Neuroscience 2021;5:434–50. https://doi.org/10.1162/netn_a_00184.

[45]	Morgan VL, Sainburg LE, Johnson GW, Janson A, Levine KK, Rogers BP, et al. Presurgical temporal lobe epilepsy connectome fingerprint for seizure outcome prediction. Brain Communications 2022;4:fcac128. https://doi.org/10.1093/braincomms/fcac128.

[46]	Bell GS, Tisi J de, Gonzalez-Fraile JC, Peacock JL, McEvoy AW, Harkness WFJ, et al. Factors affecting seizure outcome after epilepsy surgery: An observational series. Journal of Neurology, Neurosurgery & Psychiatry 2017;88:933–40. https://doi.org/10.1136/jnnp-2017-316211.

[47]	Sinha N, Dauwels J, Kaiser M, Cash SS, Brandon Westover M, Wang Y, et al. Predicting neurosurgical outcomes in focal epilepsy patients using computational modelling. Brain 2017;140:319–32. https://doi.org/10.1093/brain/aww299.

[48]     Kuroda N, Sonoda M, Miyakoshi M, Nariai H, Jeong J-W, Motoi H, et al. Objective interictal electrophysiology biomarkers optimize prediction of epilepsy surgery outcome. Brain Communications 2021;3:fcab042. https://doi.org/10.1093/braincomms/fcab042.

[49]     Memarian N, Kim S, Dewar S, Engel J, Staba RJ. Multimodal data and machine learning for surgery outcome prediction in complicated cases of mesial temporal lobe epilepsy. Computers in Biology and Medicine 2015;64:67–78. https://doi.org/10.1016/j.compbiomed.2015.06.008.

[50]     Bernabei JM, Li A, Revell AY, Smith RJ, Gunnarsdottir KM, Ong IZ, et al. Quantitative approaches to guide epilepsy surgery from intracranial EEG. Brain 2023:awad007. https://doi.org/10.1093/brain/awad007.